\pdfoutput=1

\documentclass[pra,twocolumn,10pt,superscriptaddress, footnoteinbib]{revtex4}
\usepackage{amsmath}
\usepackage{latexsym}
\usepackage{amssymb}
\usepackage{graphics,epstopdf}
\usepackage[colorlinks=true, citecolor=blue, urlcolor=blue ]{hyperref}
\usepackage{epsf,graphics,graphicx}

\newcommand{\n}{\noindent}

\newcommand{\ed}{\end{document}}
\newcommand{\beq}{\begin{equation}}
\newcommand{\eeq}{\end{equation}}

\begin{document}
\title{Unified approach towards the dynamics of optical and electron vortex beams}
\author{ Pratul Bandyopadhyay \footnote{Electronic address:{b$_{-}$pratul@yahoo.co.in}}, Banasri Basu\footnote{Electronic address:{sribbasu@gmail.com}}${}^{}$, Debashree Chowdhury\footnote{Electronic
address:{debashreephys@gmail.com}, Present affiliation: Department of Physics, Indian Institute of Technology, Kanpur 208016, India}${}^{}$,  }
\affiliation{Physics and
Applied Mathematics Unit, Indian Statistical Institute,\\
 203,Barrackpore Trunk Road, Kolkata 700 108, India}


\begin{abstract}
\n
We have proposed  a unified framework towards the dynamics of optical and electron vortex beams from the perspective of geometric phase and the associated Hall effects. The unification is attributed to the notion that the spin degrees of freedom of a relativistic particle, either massive or massless, is associated with a vortex line. 
Based on cylindrical corodinate formulation, which leads to a local vortex structure related to orbital angular momentum (OAM), it can be shown that when electron vortex beams (EVBs) move in an external electric field paraxial beams give rise to OAM Hall effect and  non-paraxial beams with tilted vortices initiate spin Hall effect in free space as well as in an external field.
A similar analysis reveals that the paraxial optical vortex beams(OVBs) in an inhomogeneous medium induce OAM Hall effect whereas non-paraxial beams with tilted vortices drive the spin Hall effect. Moreover, both OVBs and EVBs with tilted vortices give rise to OAM states with  arbitrary fractional value.

\end{abstract}

\maketitle

 Research into the properties of vortices has thrived since a remarkable paper of Nye and Berry \cite{1}  described the  basic properties of $dislocations$ $in$ $wave trains.$ They have shown that in 3D space optical vortex beams (OVBs) in general contain dislocation lines when the phase becomes singular and currents coil around the vortex line. When the vortex line is parallel(orthogonal) to the wave propagation direction, this is characterized as screw(edge) dislocation. Also there may be mixed screw-edge dislocations, which are tilted with respect to the propagation direction. 
Subsequently, it has been shown that the optical beams with screw dislocations possess quantized orbital angular momentum (OAM) directed along the beam axis \cite{2} and  OVBs with mixed edge-screw dislocations with tilted vortices carry OAM in an arbitrary direction \cite{3}. Exact solution of the Helmholtz equation can be constructed possessing wave front dislocation lines in the form of loops, links and knots when we have a superposition of the Bessel beams \cite{7}
Since the Helmholtz equation corresponds to the stationary situation of the the Schr\"odinger equation in free space, OVBs and non- relativistic electron vortex beams (EVBs) are expected to share similar geometric properties. Indeed, the existence of EVBs in free space for non relativistic electrons.
has been predicted in \cite{4}.
Later, EVBs  carrying OAM have been produced experimentally \cite{5,verbeek} 
 and in the study of relativistic EVBs,  the exact Bessel beam solutions of the Dirac equation have been derived \cite{6}.

Recently, we  have pointed out  \cite{9} that in the Skyrmionic model of an electron, where it is depicted as a scalar electron moving around a vortex line which is topologically equivalent to a magnetic flux line giving rise to the spin degrees of freedom, EVBs appear as a natural consequence. The geometrodynamics of the vortex beams has been analysed from the perspective of the geometric phase (Berry phase) \cite{berry}.  
When the Berry phase acquired by the scalar electron encircling the vortex line involves quantized Dirac monopole, we have paraxial(non-paraxial) beam with the vortex line parallel(orthogonal) to the wavefront propagation direction (z-axis). If the vortex line is tilted with respect to the propagation direction the Berry phase involves non-quantized monopole. Bessel beams in the paraxial case correspond to the situation when the polar angle of the plane wave wave vector with 
the $z$ axis $\theta_{0}\longrightarrow 0.$ For non-paraxial beams with tilted vortices Bessel beams correspond to the situation when the polar angle $\theta_{0}$ 
takes an arbitrary  non-zero value. Within this framework, we have investigated the propagation of electron vortex beams in a magnetic field \cite{pra15}  and have also theoretically studied the orbital angular momentum (OAM) Hall effect and spin Hall effect of EVBs interacting with a  laser field \cite{prl15}. Our motivation here is to 
propose a unified framework towards the dynamics of OVBs and EVBs from the perspective of the geometric phase.



To this end, we start with the note that 
due to the presence of the spin vector, the localization region of a relativistic massless as well as massive particle is found to be $S^3$ \cite{10,11}. Since $S^{3}$ is equivalent to SU(2) and  $S^{2}=\frac{SU(2)}{U(1)},$  the sphere 
$S^{3}$ can be constructed from the 3D compact space $S^{2}$ by Hopf fibration. The Abelian field U(1), which causes the Hopf fibration, corresponds to a monopole bundle.  
The monopole bundle gives rise to the spin of the system and the magnetic flux line associated with this effectively represents a vortex line. For an electron, the monopole strength $|\mu| = \frac{1}{2}$ represents the spin of the electron. In fact, the total angular momentum of a charged particle in the field of a monopole is given by $\vec{J} = \vec{L} - \mu\hat{\vec{r}},$ where $\vec{L}$ is the OAM and $\mu$ is the monopole charge. When OAM is zero, the total angular momentum of the system, which is effectively the spin is given by $|\mu|,$ with $s_{z} = \pm \mu.$ In this context it may be added that spin degrees of freedom are represented as SU(2) gauge bundle. The general property of a non-Abelian gauge theory is that when the topological $\theta$-term is introduced in the Lagrangian, a topologically nontrivial Abelian gauge field corresponding to a vortex line (magnetic flux line) arises in the configuration space \cite{12,13}. The situation essentially gives rise to a loop space when a vortex line is enclosed by a loop. This depicts a fermion as a scalar particle encircling a magnetic flux line representing the spin degrees of freedom, which may be viewed as a Skyrmion. In this framework, the vortex associated to the U(1) field corresponding to Hopf fibration may be considered as a
 spin vortex. For a vector particle, such as a photon, we can treat it as a composite of two spin $\frac{1}{2}$ states and the spin is given by $2\mu$ with $\mu = \frac{1}{2}.$ Indeed, for an anisotropic spin system with spin $S\ge 1,$ the effect of the Dirac string can be avoided if we consider the spin system as a correlated system of composite one-half spins \cite{14}. Taking advantage of this formulation, we  generalize the result for photon, a  massless spin $1$ particle.  

The Berry phase acquired by a quantum particle, while traversing a closed path encircling magnetic  flux lines is given by $2\pi\mu,$ $\mu$ being the corresponding monopole charge\cite{16}. When the monopole is located at the origin of a unit sphere, the Berry phase is given by $\mu\Omega(C)$ where $\Omega(C)$ is the solid angle subtended by the closed contour at the origin and is given by \beq \Omega(C) = \int_{C}(1 - cos \theta)d\phi = 2\pi(1- cos\theta),\label{6}\eeq where $\theta$ is the polar angle of the vortex line measured from  the quantization axis(z axis).  
	In case of OVBs (EVBs) the beam field (scalar electron) orbiting around the vortex line gives rise to this geometric phase (Berry phase) . When the beam field (scalar electron) rotating around the vortex line acquires certain quantized OAM $l$, the corresponding total phase is given by $2\pi l+\phi_B$, $\phi_B$ being the Berry phase. For quantized OAM with $l\in Z$, the effective phase for such a system is $\phi_B$ as the factor $2\pi l$ leads to a trivial phase.
The Berry phase acquired by the beam field moving around the vortex line in an OVB is thus given by $ \phi_{B(o)} = 2\pi(1 - cos\theta)\label{7},$ as $\mu = 1,$ corresponds to the photon spin. The phase acquired by the scalar electron moving around the vortex line in an EVB, where the relevant $\mu = \frac{1}{2},$ is given by $ \phi_{B(e)} = \pi(1 - cos\theta)\label{8}.$ 
      Transforming to the reference frame where the beam field (scalar electron) is considered to be fixed and the vortex state (spin state) moves in the field of a magnetic monopole around a closed path, $\phi_{B(o)}(\phi_{B(e)})$ corresponds to the geometric phase acquired by the vortex state. The angle $\theta$ corresponds  to the deviation of the vortex line from the $z$ axis. Equating $\phi_{B(o)}(\phi_{B(e)})$ with $2\pi\mu,$ we find that the effective monopole charge associated with the vortex line in an OVB(EVB) with polar angle $\theta$ measured from  the $z$ axis is given by
     \beq \mu_{(o)}\left( \mu_{(e)}\right) = (1-cos\theta)\left(\frac{1}{2}(1 - cos\theta)\right)\label{9}.\eeq
     Indeed, when $\theta = 0,$ we have paraxial beams and correspond to screw dislocation, whereas for $\theta =  \frac{\pi}{2},$ we have non-paraxial beams corresponding to edge dislocation. However, for an arbitrary angle $0 < ~\theta<~\frac{\pi}{2},$  we have non-paraxial beams which correspond to mixed screw-edge dislocation. In the latter case, these correspond to tilted vortices.
    In both the cases of OVBs and EVBs, the corresponding monopole charge $\mu$ is quantized for $\theta = 0$ and $\frac{\pi}{2}$, whereas for any arbitrary angle $\theta,$ with $0 < ~\theta<~\frac{\pi}{2},$ the monopole charge is non-quantized.

For OVBs and EVBs, vortices are phase singularities which appear when the corresponding waves are superposed.  In cylindrical coordinate formulation, a local vortex structure appears which is associated with the OAM. 
In case of EVBs, the solution of Dirac equation in cylindrical coordinates gives rise to Bessel beams \cite{6}.

 The Dirac equation is given by(c=$\hbar$=1)
\beq i\partial_{t}\psi = \left(\vec{\alpha} . \vec {p} + \beta m\right)\psi \label{4},\eeq where $\vec{\alpha}$ and $\beta$ are the $4\times 4$ Dirac matrices, $\vec{p} = -i\partial_{\vec{r}}$ is the momentum operator and $m$ is the electron mass. 
The solution for the plane waves can be obtained using cylindrical coordinates $(\rho, \varphi, z)$ in real space and $(p_{\perp} , \phi, p_{\parallel}) = (p sin\theta, \phi, p cos\theta)$ in momentum space.
	Assuming that the polarization amplitudes are the same for all the plane waves, we obtain the solution \cite{6} as 
   \begin{widetext}
   \beq \psi_{l} = \frac{exp i\Phi}{\sqrt 2}\left[\left(\begin{array}{cr}
      \sqrt{1 + \frac{m}{E}}w \\
       \sqrt{1 - \frac{m}{E}}\sigma_{z}cos\theta_{0}w
   \end{array}\right)e^{il\varphi}J_{l}(\xi) + i \left(\begin{array}{cccr}
        0  \\
       0 \\
    -\beta\sqrt{\Delta}\\
          0
   
   \end{array}\right)e^{i(l -1)\varphi}J_{l-1}(\xi) +i \left(\begin{array}{cccr}
              0  \\
                0 \\
                0 \\
   \alpha\sqrt{\Delta}
              \end{array}\right)e^{i(l +1)\varphi}J_{l+1}(\xi)\right] \label{18}\eeq
              \end{widetext}
   where $\Delta = (1-\frac{m}{E})sin^{2}\theta_{0}.$  Here,  $p_{\perp 0} = p sin\theta_{0},$ $p_{\parallel 0} = p cos\theta_{0},$ $\theta_{0}$ being the fixed polar angle with the $z$ axis and  $\Phi = (p_{\parallel 0}z - Et)$ and $\xi = p_{\perp 0}r.$  The first term in the square bracket represents a Bessel beam of the order of $l$, $\psi_{l} \propto J_{l}(\xi)e^{i(l\varphi + \Phi)}. $ The terms proportional to $\sqrt{\Delta}$ represent the spin-orbit interaction( SOI) term.
The local vortex structure $e^{il\phi}$ in the wave packet is associated with a magnetic-monopole-type gauge field contributing to the Berry connection. Indeed, considering a unit vector $\vec{e}$ orthogonal to $\vec{p},$ so that under variation of $\vec{p},$ $\vec{e}$ moves in the unit sphere $\frac{\vec{p}}{p},$ a magnetic-monopole-type connection $\vec{{\cal A}}$ is generated through the variation of $\vec{p}$ as
   \beq \vec{{\cal A}} = \left(i (e_{x} - ie_{y})\frac{\partial}{\partial \vec{p}}(e_{x} + ie_{y})\right)\label{5}.\eeq
   Noting that $e^{il\phi} =  \left(e_{x} + ie_{y}\right)^{l},$ we can write the OAM dependent gauge field as $\vec{{\cal A}}^{(l)} = l\vec{{\cal A}}.$

For monochromatic paraxial electromagnetic beam  propagating in a smoothly inhomogeneous isotropic medium with definite values of spin angular momentum (SAM) and OAM, the beam's electric field can be described (in local cylindrical coordinate ($\rho,\phi,z$)) as \cite{15}
\beq E^{\rho,l,s} = \vec{e}^{~s}F^{p,|l|}(\rho)exp[il\phi+i\int k dz]\label{1}.\eeq Here, $k(z)$ is the central wave vector directed along the beam axis $z,$ $\vec{e}^{~s}$ is the unit vector depicting the polarization of wave with the helicity $s_{z} = \pm 1$ and $F^{p,|l|}$ is the radial function with quantum number $p = 0, 1,2,....n$ and $l$ is the azimuthal quantum number with $l = 0,\pm1,\pm 2.... \pm p$ which is the value of the intrinsic orbital angular momentum. The expression is valid in the diffractionless approximation. The variation of the ray direction $\vec{k}$ gives rise to a parallel transport law. Indeed, the spin related Berry connection in momentum space associated with the monopole may be represented as 
\beq \vec{{\cal A}}^{(s)} = i\vec{e}^{~s^{\dagger}}\frac{\partial}{\partial\vec{k}}\vec{e}^{~s}, \label{2}\eeq     
with $s=2\mu$ and $\mu = \frac{1}{2}.$
The Berry connection is associated with the monopole bundle which essentially gives rise to the spin degrees of freedom.
 However, for OAM, the Berry connection is also of  magnetic monopole type and is realized when we take resort to the local cylindrical coordinates.The OAM related Berry connection is
represented as 
\beq \vec{{\cal A}}^{(l)} = i \int_0^{2\pi} 
e^{-il\phi}\frac{\partial}{\partial\vec{k}}e^{il\phi} = l\vec{{\cal A}}, \label{3}\eeq

For paraxial beams in both EVBs and OVBs as $\theta = 0,$ we have vanishing Berry phase and there is no spin orbit interaction(SOI). There is  OAM Hall effect due to the Berry connection associated with the local vortex structure $e^{il\phi}.$ Indeed, for EVBs in an external electric field, electrons are accelerated and the anomalous velocity caused by the Berry curvature associated with the connection $\vec{{\cal A}}^{(l)}$ is given by
 \beq \vec{v} =\dot {\vec{p}}\times \vec{\Omega}^{(l)}(\vec{p}) = l\dot {\vec{p}}\times \frac{\vec{p}}{p^{3}}.\label{15}\eeq  
Here, dot denotes the derivative with respect to time.
 This leads to the OAM Hall effect in EVBs. In OAM Hall effect we have splitting of beams with opposite values of $l$ giving rise to the transverse current of OAM.  
   A similar situation arises for paraxial beams (monochromatic vortex waves) in OVBs in a smooth inhomogeneous medium when the Berry curvature is related to the connection $\vec{{\cal A}}^{(l)}.$ In an inhomogeneous medium with refractive index $n(r),$ we have
 \beq \dot{\vec{k}} = k\vec{\nabla} ln~n, \dot{\vec{r}} = \frac{\vec{k}}{|k|} +\dot{\vec{k}} \times \vec{\Omega}^{l}(k),\eeq with $\vec{\Omega}^{l}(k) = l\frac{\vec{k}}{k^{3}}.$ Here, dot denotes the derivative with respect to $z.$ This leads to the ray deflection $\delta \vec{r} = \int \vec{\Omega}^{l}(\vec{k})\times d\vec{k}$ in the paraxial case and corresponds to OAM Hall effect.
However, for OVBs and EVBs in paraxial regime, there is no contribution from the spin related Berry curvature. This is evident from the expressions of $ \phi_{B(o)}$ and $ \phi_{B(e)}$ that show the vanishing of the relevant Berry phases for  both the cases.  
 
 The non-paraxial OVBs and EVBs with tilted vortices involve spin orbit interaction (SOI) and non-quantized monopole charge that realizes the spin Hall effect.
It is known  \cite{14,17} that  the monopole charge $\mu,$ dependent on a certain parameter $\lambda,$  in $3+1$ dimensions undergoes the renormalization group (RG) flow with the properties: \\
 - $\mu$ is stationary at some fixed points $\lambda^{*}$ of the RG flow, where $\mu(\lambda^{*})$ is equal to the monopole charge $\mu$ having quantized values $0$, $\pm\frac{1}{2},$ $\pm 1$....\\
 -$\mu$ decreases along the RG flow as $L\frac{\partial \mu}{\partial L}\leq 0,$ where $L$ is a length scale \cite{14,17}. \\
Transforming  the length scale to the time scale by $L = ct,$ a time dependent monopole charge $\tilde{\mu}(t)$ is accomplished that is non-quantized. Both EVBs and OVBs with tilted vortices are characterized by $\tilde{\mu}(t),$ which is the monopole charge associated with the corresponding vortex line. 

In case of EVBs, time dependence of the associated gauge field generates an electric field. As a result  the electrons are accelerated giving rise to an anomalous velocity. Indeed, we can introduce a non-inertial coordinate frame with basis vector $(\vec{f}, \vec{w}, \vec{u}),$ attached to the local direction of momentum $\vec{u} = \frac{\vec{p}}{p}.$ This coordinate frame rotates as $\vec{p}$ varies with time. Such rotation with respect to a motionless (laboratory) coordinate frame describes a precession of the triad $(\vec{f}, \vec{w}, \vec{u}),$ with some angular velocity. Let us restrict the direction of the vortex line at any instant of time as the local $z$ axis representing  the direction of propagation of the wave front. This corresponds to the paraxial beam in the local frame such that the local value of $\tilde{\mu}$ is changed and attains  the quantized value $\mu = \frac{1}{2},$ which follows from the precession of the spin vector. In the local non-inertial frame this corresponds to the pseudospin. The pseudospin vector is parallel to the momentum vector $\vec{p}.$ Now an anomalous velocity $\vec{v}_{a}$ will arise due to the Berry curvature $\vec{\Omega}^s(\vec{p}) = \mu\frac{\vec{k}}{p^{3}}$  and we write
 \beq \vec{v}_{a} =  \mu \dot{\vec{p}}\times\frac{\vec{p}}{p^3}. \label{41}\eeq
 The anomalous velocity is perpendicular to the pseudospin vector and points along the opposite directions depending on the chirality  $s_{z} = +\frac{1}{2}(-\frac{1}{2})$ corresponding to $\mu> 0(< 0).$ This gives rise to spin Hall effect. 
 $\vec{v}_{a}$ in eqn (\ref{41}) can be rewritten in terms of the unit vector $\vec{u} = \frac{\vec{p}}{p}$ and its time derivative $\dot{\vec{u}}$ as  
$ \vec{v}_{a} =  \mu \dot{\vec{u}}\times \vec{u}. \label{19}$
Denoting $\frac{\dot{\vec{u}}}{|\dot{\vec{u}}|} = \vec{n},$ we note that the spin current is orthogonal to the local plane ($\vec{u}, \vec{n}$) and the spin Hall effect can be understood as a Coriolis type transverse deflection and represents a real deflection of the wave trajectory \cite{9}. 

A similar situation arises for OVBs in a smooth in-homogeneous medium. The OVBs with tilted vortices give rise to spin Hall effect of light or optical Magnus effect \cite{18}.
 In an inhomogenious medium,  with refractive index $n({\bf r}),$ we have
\beq \dot{\vec{k}} = k\vec{\nabla} ln~n, \dot{\vec{r}} = \frac{\vec{k}}{k} +\dot{\vec{k}} \times \vec{\Omega}^{s}(k),\eeq with $\vec{\Omega}^{s}(\vec{k}) = (2\mu)\frac{\vec{k}}{k^{3}}.$ Here, dot denotes the derivative with respect to $z.$ $\vec{\Omega}^{s}({\vec k})$ is the Berry curvature  $(2\mu)\frac{\vec{k}}{k^{3}},$ $2\mu$ corresponding to the helicity $+1$ or $-1$ depending on $\mu> 0(\mu<0).$ The second term in the equation for $\dot{\vec{r}}$ corresponds to the transverse deflection which gives rise to spin Hall effect. This essentially leads to the ray deflection given by $\delta \vec{r} = \int \vec{\Omega}^{s}(\vec{k}) \times d\vec{k}$ in the non-paraxial case.

This spin Hall effect basically corresponds to a polarization dependent trajectory perturbation.
 Since, for the non-paraxial beams with tilted vortices the OAM vector is non-collinear with the momentum vector \cite{3}, there will be no contribution from the OAM related Berry curvature $\vec{\Omega}^{l}$ for both the cases of OVBs and EVBs. 

The SOI involved in the non-paraxial beams modifies the OAM ($\vec{L}$)and spin angular momentum (SAM) ($\vec{S}$) of the EVBs and OVBs as well. Introducing the mapping $ \vec{L}\rightarrow l\hat{\vec{z}}, ~~~\vec{S}\rightarrow s\hat{\vec{z}} ,$ we note that for EVBs with tilted vortices which incorporate non-quantized monopole charge $\tilde{\mu},$ the expectation values of  $\vec{\tilde{L}}$ and $\vec{\tilde{S}}$ are modified as \cite{9} 
\begin{eqnarray}\label{100}
 \langle\vec{\tilde{L}}\rangle = (l - \tilde{\mu})\hat{\vec{z}}~~~~~
 \langle\vec{\tilde{S}}\rangle = (s + \tilde{\mu})\hat{\vec{z}},
 \end{eqnarray}
 which follows from the relation $\vec{\tilde{L}} + \vec{\tilde{S}} = \vec{L} + \vec{S}$. Thus for EVBs with tilted vortices just like a pseudospin $\vec{\tilde{S}},$  an OAM is generated denoted  by $\vec{\tilde{L}},$ which can take any arbitrary fractional value.
The third component of the pseudospin can take any arbitrary value between $-\frac{1}{2}$ and $+\frac{1}{2}$ and as such for fractional OAM, apart from the integer part $l$, the fractional component $\tilde{\mu}$ lies between $0$ and $1$.

 For OVBs, the expectation value of the spin vector is given by 
$ \left\langle \vec{S}\right\rangle = \langle\psi|\vec{\sigma}|\psi\rangle$ and  $\vec{S}^{2} = 1\label{20}.$ For tilted vortices the SOI will induce a change  in the expectation value of spin angular momentum (SAM) as well as of OAM and  the modified  pseudospin  $\langle \tilde{\vec{S}}\rangle$ and OAM $\langle \tilde{\vec{L}}\rangle$ are given by 
 \beq \langle \tilde{\vec{S}}\rangle = (s-2\tilde{\mu})~~, \langle \tilde{\vec{L}}\rangle = (l+2\tilde{\mu}).\eeq
 It is noted that the third component of the pseudospin can take any arbitrary value between $+1$ and $-1.$ The OAM can take any arbitrary fractional value such that apart from the integer part, the fractional component lies between $0$ and $1$. 
 
A quantum mechanical formulation of fractional OAM has already  been developed \cite{21}. Unlike the integer OAM states, the fractional OAM states require the introduction of an additional parameter denoting the orientation of the phase discontinuity. In fact, as these states involve non-quantized monopoles, the wave function becomes multivalued. This can be rendered a single valued function by introducing  a branch cut, the position of which is taken to be the position of the discontinuity \cite{21}. The functional time dependence of the non quantized monopole charge,  which is evident from the RG flow, implies that the beams with fractional OAM will be unstable on propagation.  The stability can be attained if the effect of the Gouy phase associated with the superposition of Laguerre-Gaussian modes generating OVBs can be regulated \cite{21}. A similar situation may be  developed for the propagation of EVBs in an external magnetic field \cite {pra15}. 
The vortex structure of light emerging from a fractional  phase step is characterised by a chain  of alternating vortices such that every pair of it forms  a hairpin when the vortices converge at a common turning point \cite{19, 20}. 
In the present formalism, the fractional OAM is generated when the relevant Berry phase involves non quantized monopole and we have to consider the effect of the Dirac string. Indeed, this requires the introduction of a topological term in the Berry phase apart from the usual geometric component \cite{bruno}. However, the effect of Dirac string can be avoided when we consider the entangled pairs of spins in a spin system \cite{14}.  Since the vortices in OVBs and EVBs are considered as spin vortices, entangled pairs of vortices may be viewed as entangled pair of spins. The dynamics of the EVBs and OVBs can be studied as a model of a spin system and the entangled pair of spins may represent a pair of entangled vortices.  One can easily argue that the entangled pair of tilted vortices involve alternating vortices and forms the structure of a hairpin. Conclusively, the vortex structure of OVBs and EVBs with fractional OAM  can thus be characterized by a chain of pairs of alternating vortices forming the shape of hairpins.  
The study of  fractional OAM in the context of two photon entanglement \cite{25} may inspire one to study the similar features in two electron entangled states.

The advancement of experiments \cite{yao2010} on vortex beams and their  wide range of applications \cite{verbeek,yao2010,molline, kaplars} mark  the significance of a unified theory for the dynamics of OVBs and EVBs.  
The vortex beams carrying OAM can be produced by spiral phase  plates and various other devices \cite{yao2010}.  
The spiral phase plates with noninteger phase steps \cite{opticcomon} as well as special holograms \cite{jetp} have been used to generate light beams with fractional OAM. Moreover, spatial light modulators  or lasers emitting helical Lauguerre-Gaussian modes can be used to produce OVBs \cite{verbeek}.  The OAM carried by the optical beam field helps it to trap and rotate colloid particles, living cells and to act as $optical $ $spanner$ which can be used in various applications \cite{molline, kaplars}. Individual photons carrying  OAM makes it possible to be used in the theory of quantum information  processing  (QIP) \cite{molline}. EVBs can be produced by making versatile holographic reconstruction technique in a transmission electron microscope \cite{verbeek}. Electron vortices are used for the magnetic state mapping \cite{verbeek}. This key idea becomes very useful for condensed matter physics and material science as it bridges atomic scale resolution with the local magnetization inside samples and also plays a decisive role in the context of spinronics device applications.
 Both OVBs and EVBs have been used to observe the effect of dynamic electron correlations in interaction of atoms and molecules \cite{kaplars}.  
 
Recently, apart from the EVBs and OVBs, acoustic vortex beams  have been studied in paraxial and non-paraxial regimes  \cite{marston99,  marchianoprl,thomaspre10, marstonpre11, bliokhprb}. Acoustic vortex  beams (AVBs)  
 with  an  extra  azimuthal  phase  dependence, carry orbital angular momentum and  have a helicoidal wavefront \cite{marston99,marstonpre11}.  Similarity  between the linearized elasticity equations and Maxwell equations facilitates to map polarization phenomena in optics into transverse acoustic waves \cite{bliokhprb}.  
 AVBs  were analyzed beyond the paraxial approximation in \cite{ marstonpre11} to clarify an analogy with optical vortex beams.
The Hamiltonian of the transverse acoustic waves contains a term which can be viewed as the SOI of phonons \cite{bliokhprb}.
Alike electrons and photons,  the SOI of phonons is caused by the  Berry gauge potential  describing parallel transport in momentum space \cite{bliokhprb}.
 The Berry connection is related to the oppositely charged  magnetic monopoles located at the origin in momentum space corresponding to the phonons with opposite helicities. This leads to the transport of angular momentum of phonons analogous to the spin Hall effect. 
The indication of this analogy  encourages one to study in detail  the  dynamics of  AVBs  in the  framework of the geometric phase.    
 
In summary, we have analyzed 
the dynamics of optical and electron vortex beams from the viewpoint of the  geometric phase in a unified way where the OAM Hall effect and spin Hall effect play a crucial role. 


\begin{thebibliography}{999}

\bibitem{1} J. F. Nye  and M. V. Berry, Proc. R. Soc.(London) A. {\bf 336}, 165, (1974).
\bibitem{2} L. Allen,  M. W. Beijersbergen , R. J. C. Spreeuw and   J. P. Woerdman, Phys. Rev. A {\bf 45}, 8185, (1992).
\bibitem{3} K. Y. Bliokh, F. Nori, Phys. Rev.A, {\bf 86}, 033824, (2012).
\bibitem{7}  M. V. Berry and M.R. Dennis, Proc. R. Soc. (London) A, {\bf 457}, 2251, (2001), J. Leach, M.R. Dennis and M.J. Padget, Nature {\bf 432}, 165 (2004),L. Leach, M.R. Dennis, J. Courtial and M.J. Padget, New J. Phys. {\bf 7}, 55 (2005).
\bibitem{4}  K. Y. Bliokh et. al. Phys.Rev.Lett, {\bf 99}, 190404 (2007 ).
 \bibitem{5} M. Uchida and A. Tonomura,  Nature {\bf 464}, 737, (2010);
 B. J. McMorran et. al  Science, {\bf 331}, 192, (2011).
\bibitem{verbeek} J. Verbeeck, H.Tian and P. Schattschneider, Nature {\bf 467}, 301, (2010).
\bibitem{6}  K. Y. Bliokh et. al. Phys.Rev.Lett, {\bf 107}, 174802, (2011).
\bibitem{9}  P. Bandyopadhyay, B. Basu, D. Chowdhury, Proc.R.Soc.(London) A {\bf 470}, 20130525, (2014).
\bibitem{berry} M. V. Berry, Proc. R. Soc. (London) A, {\bf 392}, 45, (1984)
\bibitem{pra15}  D. Chowdhury, B. Basu, P. Bandyopadhyay, Phys. Rev. A {\bf 91}, 033812 (2015).
\bibitem{prl15} P. Bandyopadhyay, B. Basu, D. Chowdhury, Phys. Rev. Lett. {\bf 115}, 194801 (2015).
\bibitem{10}F Bayen and J. Niederlie, Czech. J. Phys. B {\bf 31}, 1317 (1981)
\bibitem{11}P. Bandyopadhyay,Int. J. Theor. Phys, {\bf 26}, 131, (1987); {\it Geometry, Topology and quantization}, Kluwer Academic publishers (1996).
.\bibitem{12} Y.S. Wu and A. Zee,  Nucl. Phys. B {\bf 258}, 157, (1985).
\bibitem{13} K. Sen, P. Bandyopadhyay,  J. Math. Phys. {\bf 35}, 2270, (1994).
\bibitem{14} P. Bandyopadhyay, Proc. R. Soc. (London) A, {\bf 467}, 427, (2011).
\bibitem{16} D. Banerjee, , P. Bandyopadhyay, J. Math. Phys. {\bf 33}, 990, (1992).
\bibitem{15} K. Y. Bliokh, Phys.Rev.Lett. {\bf 97}, 043901, (2006).
\bibitem{17}P. Bandyopadhyay, Int. J. Mod Phys. A {\bf 15}, 1415, (2000); 
\bibitem{18}K. Y.Bliokh, J. Opt. A: Pure Appl. Opt. {\bf 11}, 094009, (2009).
\bibitem{21} J.B.Gotte et al., J.Mod.Opt. {\bf 54}, 1723, (2006), J.B.Gotte et al.,Optics Express {\bf 16}, 993,, (2008).
\bibitem{19}M.V Berry, J. Opt. A: Pure Appl. Opt., {\bf 6}, 259, (2004).
\bibitem{20} J. I. Leach, F. Yao and M. J. Padgett, New J. Phys. {\bf 6}, 71 (2004); K. O'Holleran, M. R Dennis and M. J Padgett,J. European Opt. Society(R) {\bf 1}, 06008,(2006). 
\bibitem{bruno} P. Bruno, Phys. Rev. Lett. {\bf 93}, 247202 (2004)
\bibitem{25} S.S.R. Oemrawsingh et. al. Phys. Rev.Lett {\bf 92}, 217901,(2004), G.F. Calvo et. al. Phys. Rev. A, {\bf 75}, 012319, (2007).
\bibitem{yao2010} F.Yao and M.J. Padgett, Adv in Optics and Photonics, {\bf 3},161 (2011).
\bibitem{molline}  Molina Terriza et. al, Nature Phys,{\bf 3}, 305 (2007).
\bibitem{kaplars} L. Kaplan and J.H. McGuire, Phys Rev A, {\bf 92},
032702 (2015). 
\bibitem{opticcomon} M. W. Beijersbergen et al. Opt. Commun.{\bf 112}, 34 (1994).
\bibitem{jetp} V. Y. Bashenov et al. JETP Lett. {\bf 52} 429 (1990)
\bibitem{marston99} B. T. Hefner  and P. L. Marston  J. Acoust. Soc. Am. {\bf 106},
3313(1999).
\bibitem{marchianoprl}Christine E. M. Demore et. al., Phys. Rev. Lett. {\bf 108}, 194301(2012).
\bibitem{thomaspre10}D. Baresch, J.-L. Thomas and R. Marchiano, PRL {\bf 116}, 024301, (2016).
\bibitem{marstonpre11} L. Zhang and P. L. Marston, Phys. Rev. E {\bf 84}, 065601(R) (2011)
\bibitem{bliokhprb} K.Yu. Bliokh and V.D. Freilikher, Phys.Rev.B {\bf 74}, 174302,(2006).

\end{thebibliography}
\end{document}